\documentclass[copyright]{eptcs}
\providecommand{\event}{DSL 2011} 
\usepackage{breakurl}             

\usepackage[all]{xy}
\usepackage{amsfonts}
\usepackage{amssymb}
\usepackage{latexsym}
\usepackage{fleqn}

\newif\ifanswers\answerstrue

\def\Z{{\mathbb Z}}
\def\N{{\mathbb N}}
\let\sigmasum\sum
\let\piprod\prod

\def\keyword#1{\expandafter\def\csname#1\endcsname{\mathbf{#1}}}
\keyword{where}
\keyword{data}
\keyword{type}
\def\identifier#1{\expandafter\def\csname#1\endcsname{\mathit{#1}}}
\identifier{id}
\identifier{fmap}
\identifier{map}
\identifier{fork}
\identifier{prod}
\identifier{fst}
\identifier{snd}
\identifier{const}
\identifier{double}
\identifier{join}
\identifier{inl}
\identifier{inr}
\identifier{Stream}
\def\inn{\mathit{in}}
\identifier{fold}
\identifier{foldr}
\identifier{foldl}
\identifier{out}

\identifier{head}
\identifier{tail}
\identifier{mapS}
\identifier{unfold}
\identifier{diag}
\identifier{return}
\identifier{join}
\identifier{extr}
\identifier{dupl}
\identifier{sum}
\identifier{maximum}
\identifier{segs}
\identifier{inits}
\identifier{tails}
\identifier{scanl}
\identifier{scanr}
\identifier{scanu}
\identifier{scand}
\identifier{root}
\identifier{concat}
\identifier{return}
\identifier{opt}
\identifier{contents}

\identifier{Nothing}
\identifier{Just}
\identifier{maybe}
\def\funct#1{\mathsf{#1}}

\def\functF{\funct{F}}
\def\functG{\funct{G}}
\def\functH{\funct{H}}
\def\functK{\funct{K}}
\def\functM{\funct{M}}
\def\functL{\funct{L}}

\def\functT{\funct{T}}
\def\functU{\funct{U}}
\def\functMaybe{\funct{Maybe}}
\def\functList{\funct{List}}
\def\functTree{\funct{Tree}}
\def\functId{\funct{Id}}

\def\ntto{\mathbin{\stackrel{.}{\to}}}
\def\mplus{\mathbin{\uplus}}
\def\mzero{\emptyset}
\def\max{\mathbin{\sqcup}}
\def\append{\mathbin{{+}\!\!\!{+}}}
\def\dollar{\mathbin{\hbox{\footnotesize\$}}}

\catcode`\"=\active
\def\oq{``\let"\cq} 
\def\cq{''\let"\oq}
\let"\oq

\begin{document}

\title{Maximum Segment Sum, Monadically \\ 
  (distilled tutorial\ifanswers, with solutions\fi)}
\author{Jeremy Gibbons
\institute{Department of Computer Science, University of Oxford \\[0.3ex] Wolfson Building, Parks Road, Oxford OX1 3QD, United Kingdom}
\email{jeremy.gibbons@cs.ox.ac.uk}
}
\def\titlerunning{Maximum Segment Sum, Monadically}
\def\authorrunning{Jeremy Gibbons}

\maketitle

\begin{abstract}
  The \emph{maximum segment sum} problem is to compute, given a list
  of integers, the largest of the sums of the contiguous segments of
  that list. This problem specification maps directly onto a
  cubic-time algorithm; however, there is a very elegant linear-time
  solution too. The problem is a classic exercise in the mathematics
  of program construction, illustrating important principles such as
  calculational development, pointfree reasoning, algebraic structure,
  and datatype-genericity. Here, we take a sideways look at the
  datatype-generic version of the problem in terms of monadic
  functional programming, instead of the traditional relational
  approach; the presentation is tutorial in style, and leavened with
  exercises for the reader.
\end{abstract}

\section{Introduction} \label{sec:intro}

Domain-specific languages are one approach to the general challenge of
raising the level of abstraction in constructing software
systems. Rather than making use of the same general-purpose tools for
all domains of discourse, one identifies a particular domain of
interest, and fashions some tools specifically to embody the
abstractions relevant to that domain. The intention is that common
concerns within that domain are abstracted away within the
domain-specific tools, so that they can be dealt with once and for all
rather than being considered over and over again for each development
within the domain.

Accepting the premise of domain-specific over general-purpose tools
naturally leads to an explosion in the number of tools in the
programmer's toolbox---and consequently, greater pressure on the tool
designer, who needs powerful meta-tools to support the lightweight
design of new domain-specific abstractions for each new
domain. Language design can no longer be the preserve of large
committees and long gestation periods; it must be democratized and
streamlined, so that individual developers can aspire to toolsmithery,
crafting their own languages to address their own problems.

Perhaps the most powerful meta-tool for the aspiring toolsmith is a
programming language expressive enough to host domain-specific
\emph{embedded} languages \cite{Hudak1996:Building}. That is, rather
than designing a new domain-specific language from scratch, with
specialized syntax and a customized syntax-aware editor, a dedicated
parser, an optimization engine, an interpreter or compiler, debugging
and profiling systems, and so on, one simply writes a \emph{library}
within the host language. This can be constraining---one has to accept
the host language's syntax and semantics, which might not entirely
match the domain---but it is very lightweight, because one can exploit
all the existing infrastructure rather than having to reinvent it. 

Essentially, the requirement on the host language is that it provides
the right features for capturing new abstractions---things like
strong typing, higher-order functions, modules, classes, data
abstraction, datatype-genericity, and so on. If the toolsmith can
formalize a property of their domain, the host language should allow
them to express that formalization within the language.  One might say
that a sufficiently expressive host language is in fact a
\emph{domain-specific language for domain-specific languages}.

Given a suitably expressive host language, the toolsmith designs a
domain-specific language as a library---of combinators, or classes, or
modules, or whatever the appropriate abstraction mechanism is in that
language. Typically, this consists of a collection of
\emph{constructs} (functions, methods, datatypes) together with a
collection of \emph{laws} defining an equational theory for those
constructs. The pretty-printing libraries of Hughes
\cite{Hughes95:Design} and Wadler \cite{Wadler2003:Prettier} are a
good example; but so is the relational algebra that underlies SQL
\cite{Codd70:Relational}.

This tutorial presents an exercise in reasoning with a collection of
combinators, representative of the kinds of reasoning that can be done
with the constructs and laws of any domain-specific embedded language.
We will take Haskell \cite{PeytonJones2003:Haskell} as our host
language, since it provides many of the right features for expressing
domain-specific embedded languages. However, Haskell is still not
perfect, so we will take a somewhat abstract view of it, mixing true
Haskell syntax with some mathematical idealizations---our point is the
equational reasoning, not the language in which it is expressed. 
Section~\ref{sec:notation} provides a brief summary of our
notation, and there are some exercises 
\ifanswers with solutions \fi in Section~\ref{sec:exercises}.

\section{Maximum segment sum} \label{sec:mss}

The particular problem we will be considering is a classic exercise in
the mathematics of program construction, namely that of deriving a
linear-time algorithm for the \emph{maximum segment sum} problem,
based on \emph{Horner's Rule}. The problem was popularized in Jon
Bentley's \textit{Programming Pearls} column
\cite{Bentley84:Programming} in \textit{Communications of the ACM}
(and in the subsequent book \cite{Bentley86:Programming}), but I
learnt about it from my DPhil supervisor Richard Bird's lecture notes
on the \textit{Theory of Lists} \cite{Bird87:Theory} and
\textit{Constructive Functional Programming}
\cite{Bird88:Constructive} and his paper \textit{Algebraic Identities
  for Program Calculation} \cite{Bird89:Algebraic}, which he was
working on around the time I started my doctorate. It seems like I'm
not the only one for whom the problem is a favourite, because it has
since become a bit of a clich\'e among program calculators; but that
won't stop me revisiting it.

The original problem is as follows. Given a list of numbers (say, a
possibly empty list of integers), find the largest of the sums of the
contiguous segments of that list. In Haskell, this specification could
be written like so:
\[ \begin{array}{@{}lcl}
\mathit{mss} &::& [\mathit{Integer}] \to \mathit{Integer} \\
\mathit{mss} &=& \maximum \cdot \map\,\sum \cdot \segs
\end{array} \]
where $\segs$ computes the contiguous segments of a list:
\[ \begin{array}{@{}lcllcl}
\rlap{$\segs, \inits, \tails :: [\alpha] \to [[\alpha]]$} \\
\segs &=& \rlap{$\concat \cdot \map\,\inits \cdot \tails$} \\
\tails &=& \foldr\,f\,[[\,]]
  \quad&\where\; f\,\mathit{x}\,\mathit{xss} &=& (\mathit{x}:\head\,\mathit{xss}):\mathit{xss} \\
\inits &=& \foldr\,g\,[[\,]]
  \quad&\where\; g\,\mathit{x}\,\mathit{xss} &=& [\,] : \map\,(\mathit{x}:)\,\mathit{xss} 
\end{array} \]
and $\sum$ computes the sum of a list of integers, and $\maximum$ the maximum of a nonempty list of integers:
\[ \begin{array}{@{}lcl}
\rlap{$\sum, \maximum :: [\mathit{Integer}] \to \mathit{Integer}$} \\
\sum &=& \foldr\,(+)\,0 \\
\maximum &=& \foldr_1\,(\max)
\end{array} \]
(Here, $\max$ denotes binary maximum.)
This specification is executable, but takes cubic time; the problem is
to do better.

We can get quite a long way just using standard properties of $\map$,
$\inits$, and so on. It is straightforward (see
Exercise~\ref{ex:mss-lists}) \label{use:mss-lists} to calculate that
\[ 
\mathit{mss} = \maximum \cdot \map\,(\maximum \cdot \map\,\sum \cdot \inits) \cdot \tails
\]
If we can write $\maximum \cdot \map\,\sum \cdot
\inits$ in the form $\foldr\,h\,e$, then the $\map$ of this can be
fused with the $\tails$ to yield $\scanr\,h\,e$; this observation is
known as the \emph{Scan Lemma}. Moreover, if $h$ takes constant time,
then this gives a linear-time algorithm for $\mathit{mss}$.

The crucial observation is based on Horner's Rule for evaluation of
polynomials, which is the first important thing you learn in numerical
computing---I was literally taught it in secondary school, in my
sixth-year classes in mathematics. Here is its familiar form:
\[ \displaystyle
\sigmasum_{i=0}^{n-1} a_i x^i
   = a_0 + a_1 x + a_2 x^2 + \cdots + a_{n-1} x^{n-1}
   = a_0 + x(a_1 + x(a_2 + \cdots + x\,a_{n-1}))
\]
but the essence of the rule is about sums of products
(see Exercise~\ref{ex:horner}): \label{use:horner}
\[ \displaystyle
\sigmasum_{i=0}^{n-1} \piprod_{j=0}^{i-1} u_j
   = 1 + u_0 + u_0u_1 + \cdots + u_0u_1\ldots u_{n-2}
   = 1 + u_0(1 + u_1(1 + \cdots + u_{n-2}))
\]
Expressed in Haskell, this is captured by the equation
\[ \sum \cdot \map\,\mathit{product} \cdot \inits = \foldr\,(\oplus)\,e
  \quad \where\; e = 1 \mathbin{;} u \oplus z = e + u \times z \]
(where $\mathit{product} = \foldr\,(\times)\,1$ computes the product
of a list of integers).

But Horner's Rule is not restricted to sums and products; the essential
properties are that addition and multiplication are associative, that
multiplication has a unit, and that multiplication distributes over
addition. This the algebraic structure of a \emph{semiring} (but
without needing commutativity of addition). In particular, the so-called
\emph{tropical semiring} on the integers, in which "addition" is binary
maximum and "multiplication" is integer addition, satisfies the
requirements. So for the maximum segment sum problem, we get
\[ \maximum \cdot \map\,\sum \cdot \inits = \foldr\,(\oplus)\,e \quad
\where\; e = 0 \mathbin{;} u \oplus z = e \max (u + z) \]
Moreover, $\oplus$ takes constant time, so this gives a linear-time
algorithm for $\mathit{mss}$ (see Exercise~\ref{ex:simulate}).
\label{use:simulate} \label{use:semirings}

\section{Tail segments, datatype-generically} \label{sec:tails}

About a decade after the initial "theory of lists" work on the maximum
segment sum problem, Richard Bird, Oege de Moor, and Paul
Hoogendijk came up with a datatype-generic version of the problem 
\cite{Bird*96:Generic}. It's fairly clear what "maximum" and
"sum" mean generically, but not so clear what "segment" means for
nonlinear datatypes; the point of their paper is basically to resolve
that issue.

Recalling the definition of $\segs$ in terms of $\inits$ and $\tails$,
we see that it would suffice to develop datatype-generic notions of
"initial segment" and "tail segment". One fruitful perspective is
given in Bird \& co's paper: a "tail segment" of a cons list is just a
subterm of that list, and an "initial segment" is the list but with
some tail (that is, some subterm) replaced with the empty structure.

So, representing a generic "tail" of a data structure is easy: it's a
data structure of the same type, and a subterm of the term denoting
the original structure. A datatype-generic definition of $\tails$ is a
little trickier, though. For lists, you can see it as follows:
\emph{every node of the original list is labelled with the subterm of
  the original list rooted at that node}. I find this a helpful
observation, because it explains why the $\tails$ of a list is one
element longer than the list itself: a list with $n$ elements has
$n+1$ nodes ($n$ conses and a nil), and each of those nodes gets
labelled with one of the $n+1$ subterms of the list. Indeed, $\tails$
ought morally to take a possibly empty list and return a
\emph{non-empty list} of possibly empty lists---there are two
different datatypes involved. Similarly, if one wants the "tails" of a
data structure of a type in which some nodes have no labels (such as
leaf-labelled trees, or indeed such as the "nil" constructor of
lists), one needs a variant of the datatype providing labels at those
positions. Also, for a data structure in which some nodes have
multiple labels, or in which there are different types of labels, one
needs a variant for which \emph{every node has precisely one
  label}. 

Bird \& co call this the \emph{labelled variant} of the original
datatype; if the original is a polymorphic datatype $\functT\,\alpha =
\mu(\functF\,\alpha)$ for some binary shape functor $\functF$, then the
labelled variant is $\functL\,\alpha = \mu(\functG\,\alpha)$ where
$\functG\,\alpha\,\beta = \alpha \times \functF\,1\,\beta$---whatever
$\alpha$-labels $\functF$ may or may not have specified are ignored, and 
precisely one $\alpha$-label per node is provided. 
\label{use:labelled}
Given this insight, it is straightforward to
define a datatype-generic variant $\mathit{subterms}$ of the $\tails$
function:
\[ 
\mathit{subterms}_{\functF} 
  = \fold_{\functF}(\inn_{\functG} \cdot \fork(\inn_{\functF} \cdot \functF\,\id\,\mathit{root}, \functF\,!\,\id))
  :: \functT\,\alpha \to \functL(\functT\,\alpha)
\]
where $\mathit{root} = \fst \cdot \inn_{\functG}^{-1} =
\fold_{\functG}\,\fst :: \functL\,\alpha \to \alpha$ returns the root
label of a labelled data structure, and 
$!_{\alpha} = (\lambda a \mathbin{.} ()):: \alpha \to 1$
is the unique arrow to the unit type. (Informally, having computed the
tree of subterms for each child of a node, we make the tree of
subterms for the node itself by assembling all the child trees with the
label for this node; the label should be the whole structure rooted at
this node, which can be reconstructed from the roots of the child
trees.)  What's more, there's a datatype-generic scan lemma too:
\[ \begin{array}{@{}lcl}
\mathit{scan}_{\functF} &::& (\functF\,\alpha\,\beta \to \beta) \to \functT\,\alpha \to \functL\,\beta \\
\mathit{scan}_{\functF}\,f &=& \functL\,(\fold_{\functF}\,f) \cdot \mathit{subterms}_{\functF} \\
  &=& \fold_{\functF}(\inn_{\functG} \cdot \fork(f \cdot \functF\,\id\,\mathit{root}, \functF\,!\,\id))
\end{array} \]
(Again, the label for each node can be constructed from the root
labels of each of the child trees.)
In fact, $\mathit{subterms}$ and $\mathit{scan}$ are paramorphisms
\cite{Meertens92:Paramorphisms}, and can also be nicely written
coinductively as well as inductively \cite{Meijer*91:Functional}.
\label{use:para-unfold}

\section{Initial segments, datatype-generically} \label{sec:inits}

What about a datatype-generic "initial segment"? As suggested above,
that's obtained from the original data structure by replacing some
subterms with the empty structure. Here I think Bird \& co sell
themselves a little short, because they insist that the datatype
$\functT$ supports empty structures, which is to say, that $\functF$
is of the form $\functF\,\alpha\,\beta = 1 + \functF'\,\alpha\,\beta$
for some $\functF'$. This isn't necessary: for an arbitrary $\functF$,
we can easily manufacture the appropriate datatype $\functU$ of "data
structures in which some subterms may be replaced by empty", by
defining $\functU\,\alpha = \mu(\functH\,\alpha)$ where
$\functH\,\alpha\,\beta = 1 + \functF\,\alpha\,\beta$.

As with $\mathit{subterms}$, the datatype-generic version of $\inits$
is a bit trickier---and this time, the special case of lists is
misleading. You might think that because a list has just as many
initial segments as it does tail segments, so the labelled variant
ought to suffice just as well here too. But this doesn't work for
non-linear data structures such as trees---in general, there are many
more "initial" segments than "tail" segments (because one can make
independent choices about replacing subterms with the empty structure
in each child), and they don't align themselves conveniently with the
nodes of the original structure.

The approach I prefer here is just to use a collection
type to hold the "initial segments"; that is, a monad. This could be
the monad of finite lists, or of finite bags, or of finite sets---we
will defer until later the discussion about precisely which monad, and
write simply $\functM$. That the monad corresponds to a collection
class amounts to it supporting a "union" operator $(\mplus) ::
\functM\,\alpha \times \functM\,\alpha \to \functM\,\alpha$ for
combining two collections
(append, bag union, and set union, respectively, for lists, bags, and sets), 
and an "empty" collection $\mzero :: \functM\,\alpha$ as the unit of $\mplus$,
both of which the $\join$ of the monad should distribute over 
\cite{Lellahi&Tannen97:Calculus}:
\[ \begin{array}{@{}lcl}
\join\,\mzero &=& \mzero \\
\join\,(x \mplus y) &=& \join\,x \mplus \join\,y 
\end{array} \]
(Some authors also add the axiom 
$\join\,(\functM\,(\lambda a \mathbin{.} \mzero)\,x) = \mzero$, making 
$\mzero$ in some sense both a left and a right zero of composition.)
You can think of a computation of type $\alpha \to \functM\,\beta$ in
two equivalent ways: as a nondeterministic mapping from an $\alpha$ to
one of many---or indeed, no---possible $\beta$s, or as a deterministic 
function from an
$\alpha$ to the collection of all such $\beta$s. The choice of monad
distinguishes different flavours of nondeterminism; for example, the
finite bag monad models nondeterminism in which the multiplicity of
computations yielding the same result is significant, whereas with the
finite set monad the multiplicity is not significant.

Now we can implement the datatype-generic version of $\inits$ by
nondeterministically pruning a data structure by arbitrarily replacing
some subterms with the empty structure; or equivalently, by generating
the collection of all such prunings.
\[
\mathit{prune} 
  = \fold_{\functF}(\functM\,\inn_{\functH} \cdot \opt\,\Nothing \cdot \functM\,\Just \cdot \delta_2)
  :: \mu(\functF\,\alpha) \to \functM(\mu(\functH\,\alpha))
\]
Here, $\opt$ supplies a new alternative for a nondeterministic computation:
\[ opt\,a\,x = \return\,a \mplus x \] and
$\delta_2 :: (\functF\,\alpha)\functM \ntto \functM(\functF\alpha)$
distributes the shape functor $\functF$ over the monad $\functM$
(which can be defined for all \emph{traversable} functors
$\functF\,\alpha$---we'll say more about this in
Section~\ref{sec:distr-lists}). Informally, once you have computed all
possible ways of pruning each of the children of a node, a pruning of
the node itself is formed either as $\Just$ some node assembled from
arbitrarily pruned children, or $\Nothing$ for the empty structure.
\label{use:prune}

\section{Horner's Rule, datatype-generically} \label{sec:horner}

As we've seen, the essential property behind Horner's Rule is one of
distributivity, for example of product over sum. 
In the datatype-generic case, we will model this as
follows. We are given an $(\functF\,\alpha)$-algebra $(\beta,f)$, and
a $\functM$-algebra $(\beta,k)$; you might think of these as
"datatype-generic product" and "collection sum", respectively. Then
there are two different methods of computing a $\beta$ result from an
$\functF\,\alpha\,(\functM\,\beta)$ structure: we can either distribute
the $\functF\,\alpha$ structure over the collection(s) of $\beta$s,
compute the "product" $f$ of each structure, and then compute the "sum" $k$ of
the resulting products; or we can "sum" each collection, then compute the
"product" of the resulting structure, as illustrated in the following diagram.
\[ \xymatrix@=0.75in{
\functF\,\alpha\,(\functM\,\beta) \ar[r]^{\delta_2} \ar[d]_{\functF\,\id\,k} &
\functM(\functF\,\alpha\,\beta) \ar[r]^{\functM\,f} &
\functM\,\beta \ar[d]^{k} \\
\functF\,\alpha\,\beta \ar[rr]_{f} && 
\beta
} \]
Distributivity of "product" over
"sum" is the property that these two different methods agree.
For example, with $f :: \functF\,\N\,\N \to \N$ adding all the
naturals in an $\functF$-structure, and $k :: \functM\,\N \to \N$
finding the maximum of a collection of naturals
(returning $0$ for the empty collection), the diagram commutes 
(see Exercise~\ref{ex:distr-sum-max}). \label{use:distr-sum-max}
(To match up with the rest of the story, we have presented
distributivity in terms of a bifunctor $\functF$, although the first
parameter $\alpha$ plays no role. We could just have well have used a
unary functor, dropping the $\alpha$, and changing the distributor to
$\delta :: \functF\functM \ntto \functM\functF$.)

Note that $(\beta,k)$ is required to be an algebra for the monad
$\functM$. This means that it is not only an algebra for $\functM$ as
a functor (namely, of type $\functM\,\beta \to \beta$), but also it
should respect the extra structure of the monad: $k \cdot \return =
\id$ and $k \cdot \join = k \cdot \functM\,k$. For the special case of
monads of collections, these amount to what were called
\emph{reductions} in the old Theory of Lists \cite{Bird87:Theory} work---functions $k$
of the form $\oplus/$ for binary operator ${\oplus} :: \beta \times
\beta \to \beta$, distributing over union: $\oplus / (x \mplus y) =
(\oplus/x) \oplus (\oplus/y)$ (see Exercise~\ref{ex:binary}).
\label{use:binary} A consequence of this distributivity property
is that $\oplus$ has to satisfy all the properties that $\mplus$
does---for example, if $\mplus$ is associative, then so must $\oplus$
be, and so on, and in particular, since $\mplus$ has a unit $\mzero$,
then $\oplus$ too must have a unit $e_\oplus :: \beta$, and 
$\oplus/\mzero = e_\oplus$ is forced
(see Exercise~\ref{ex:properties}). \label{use:properties}

Recall that we modelled an "initial segment" of a structure
of type $\mu(\functF\,\alpha)$ as being of type
$\mu(\functH\,\alpha)$, where $\functH\,\alpha\,\beta = 1 +
\functF\,\alpha\,\beta$. We need to generalize "product" to work on
this extended structure, which is to say, we need to specify the value
$b$ of the "product" of the empty structure too. Then we have
$\maybe\,b\,f :: \functH\,\alpha\,\beta \to \beta$, so that
$\fold_{\functH}(\maybe\,b\,f) :: \mu(\functH\,\alpha) \to \beta$.

The datatype-generic version of Horner's Rule is then about computing
the "sum" of the "products" of each of the "initial segments" of a
data structure:
\[ {\oplus/} \cdot \functM(\fold_{\functH}(\maybe\,b\,f)) \cdot \mathit{prune} \]
We can use fold fusion to show that this composition can be computed as a single
fold, $\fold_{\functF}((b\oplus) \cdot f)$,
given the distributivity property
${\oplus/} \cdot \functM\,f \cdot \delta_2 = f \cdot \functF\,\id\,(\oplus/)$
above (see Exercise~\ref{ex:dgp-horner}). \label{use:dgp-horner}
Curiously, it doesn't seem to matter what value is chosen for $b$.

We're nearly there. We start with the traversable shape bifunctor
$\functF$, a collection monad $\functM$, and a distributive law $\delta_2 ::
(\functF\,\alpha)\functM \ntto \functM(\functF\alpha)$.  We are given
an $(\functF\,\alpha)$-algebra $(\beta,f)$, an additional element $b
:: \beta$, and a $\functM$-algebra $(\beta,{\oplus/})$, such that $f$
and $\oplus$ take constant time and $f$ distributes over $\oplus/$ in
the sense above. Then we can calculate 
(see Exercise~\ref{ex:dgp-mss}) \label{use:dgp-mss} that
\[ {\oplus/} \cdot \functM(\fold_{\functH}(\maybe\,b\,f)) \cdot \segs =
\mathord{\oplus/} \cdot \mathit{contents}_{\functL} \cdot \mathit{scan}_{\functF}((b\oplus)\cdot f)
\]
where
\[ \segs = \join \cdot \functM\,\mathit{prune} \cdot \mathit{contents}_{\functL} \cdot \mathit{subterms} :: \mu(\functF\,\alpha) \to \functM(\mu(\functH\,\alpha)) \]
and where
$\mathit{contents}_{\functL} :: \functL \ntto \functM$
computes the contents of an $\functL$-structure (which, like
$\delta_2$, can be defined using the traversability of
$\functF$).
The scan can be computed in linear time, because its body takes
constant time; moreover, the "sum" $\oplus/$ and $\mathit{contents}$
can also be computed in linear time (indeed, they can even be
fused into a single pass). 

For example, with $f :: \functF\,\Z\,\Z \to \Z$ adding all the
integers in an $\functF$-structure, $b = 0 :: \Z$, and ${\oplus} ::
\Z\times\Z \to \Z$ returning the greater of two integers, we get a
datatype-generic version of the linear-time maximum segment sum
algorithm.

\section{Distributivity reconsidered} \label{sec:distr}

There's a bit of hand-waving in Section~\ref{sec:horner} to justify
the claim that the commuting diagram there really is a kind of
distributivity. What does it have to do with the familiar
equation
$a \otimes (b \oplus c) = (a \otimes b) \oplus (a \otimes c)$
capturing distributivity of one binary operator $\otimes$ over another, $\oplus$?

Recall that $\delta_2 :: (\functF\,\alpha)\functM \ntto
\functM(\functF\,\alpha)$ distributes the shape functor $\functF$ over
the monad $\functM$ in its second argument; this is the form of
distribution over "effects" that crops up in the datatype-generic
Maximum Segment Sum problem. More generally, this works for any idiom
$\functM$; this will be important below.

Generalizing in another direction, one might think of distributing
over an idiom in both arguments of the bifunctor, via an operator $\delta : \functF \cdot (\functM \times \functM) \ntto \functM \cdot \functF$, which is
to say, $\delta_\beta :: \functF\,(\functM\beta)\,(\functM\beta) \to
\functM(\functF\beta)$, natural in the $\beta$. This is the
$\mathit{bidist}$ method of the $\mathit{Bitraversable}$ subclass of
$\mathit{Bifunctor}$ that Bruno Oliveira and I used in our
paper \cite{Gibbons&Oliveira2009:Essence} on the \textsc{Iterator} pattern;
informally, it requires just that
$\functF$ has a finite ordered sequence of "element positions". Given
$\delta$, one can define $\delta_2 = \delta \cdot
\functF\,\mathit{pure}\,\id$.

That traversability (or equivalently, distributivity over effects) for
a bifunctor $\functF$ is definable for any idiom, not just any monad,
means that one can also conveniently define an operator
$\contents_{\functH} : \functH \ntto \functList$
for any traversable unary functor $\functH$. This is because the
constant functor $\functK_{[\beta]}$ is an idiom: 
the $\mathit{pure}$ method returns the empty
list, and idiomatic application appends two lists. Then one can define
\[ \contents_{\functH} = \delta \cdot
\functH\,\mathit{wrap} \] where $\mathit{wrap}$ makes a singleton
list.  For a traversable bifunctor $\functF$, we define
$\contents_{\functF} =
\contents_{\functF\triangle}$ where $\triangle$ is the
diagonal functor; that is, $\contents_{\functF} ::
\functF\,\beta\,\beta \to [\beta]$, natural in the $\beta$.
(No constant functor is a monad, except in trivial categories, so this
convenient definition of contents doesn't work monadically. Of course,
one can use a writer monad, but this isn't quite so convenient,
because an additional step is needed to extract the output.)

One important axiom of $\delta$, suggested by Ond{\v{r}}ej
Ryp{\'a}{\v{c}}ek \cite{Rypacek2010:Labelling}, is that it should be
"natural in the contents": it should leave shape unchanged, and depend
on contents only up to the extent of their ordering. Say that a
natural transformation $\phi : \functF \ntto \functG$ between
traversable functors $\functF$ and $\functG$ "preserves contents" if
$\contents_{\functG} \cdot \phi = \contents_{\functF}$. Then, in the
case of unary functors, the formalization of "naturality in the
contents" requires $\delta$ to respect content-preserving $\phi$:
\[ \delta_{\functG} \cdot \phi = \functM\phi \cdot \delta_{\functF} : \functT\functM \ntto \functM\functG 
\quad\Leftarrow\quad
\contents_{\functG} \cdot \phi = \contents_{\functF}
\]
In particular, $\contents_{\functF} : \functF \ntto \functList$ itself preserves contents, and so we expect
\[ \delta_{\functList} \cdot \contents_{\functF} = \functM(\contents_{\functF}) \cdot \delta_{\functF} \]
to hold. 

Happily, the same generic operation $\contents_{\functF}$ provides a
datatype-generic means to "fold" over the elements of an
$\functF$-structure. Given a binary operator $\otimes ::
\beta\times\beta \to \beta$ and an initial value $b :: \beta$, we can
define an $(\functF\,\beta)$-algebra $(\beta,f)$---that is, a function
$f :: \functF\,\beta\,\beta\to\beta$---by
\[ f = \foldr\,(\otimes)\,b \cdot \contents_{\functF} \] 
This is a slight
specialization of the presentation of the datatype-generic MSS problem;
there we had $f :: \functF\,\alpha\,\beta \to
\beta$. The specialization arises because we are hoping to define such
an $f$ given a homogeneous binary operator $\otimes$. On the other
hand, the introduction of the initial value $b$ is no specialization,
as we needed such a value for the "product" of an empty "segment"
anyway.
This "generic folding" construction is just what is provided by Ross
Paterson's \textit{Data.Foldable} Haskell library \cite{Paterson2005:Data}.

\section{Reducing distributivity} \label{sec:distr-lists}

The general principle about traversals underlying Ryp{\'a}{\v{c}}ek's
paper \cite{Rypacek2010:Labelling} on labelling data structures is
that it is often helpful to reduce a general problem about traversal
over arbitrary datatypes to a more specific one about lists,
exploiting the "naturality in contents" property of traversal. We'll
use that tactic for the distributivity property in the
datatype-generic version Horner's Rule.

Consider the following diagram.
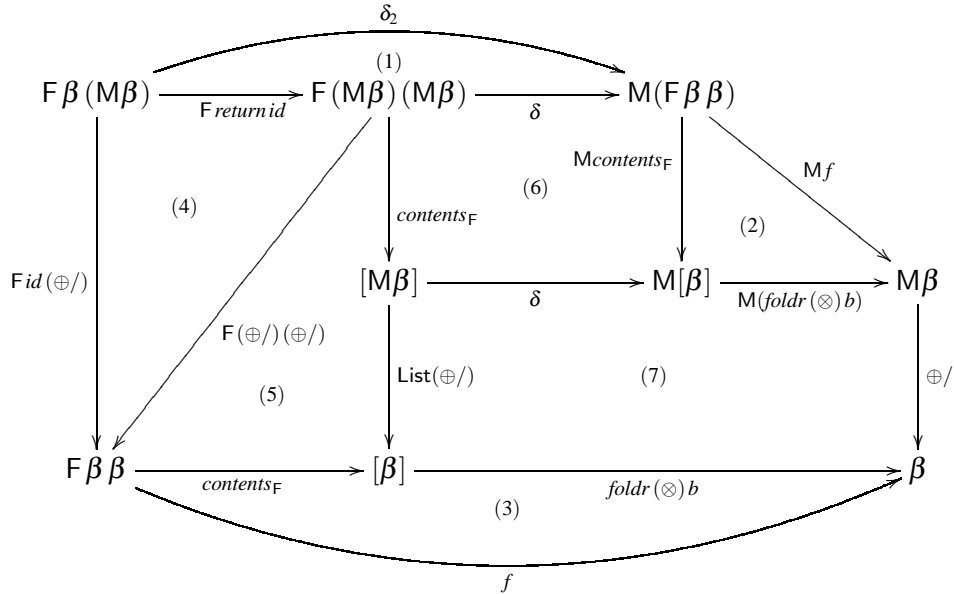
\[ \xymatrix@=0.75in{
\functF\,\beta\,(\functM\beta) 
  \ar[r]_{\functF\,\return\,\id}
  \ar[dd]_{\functF\,\id\,(\oplus/)} 
  \ar@{}[ddr]|(0.3){(4)} 
  \ar@/^2pc/[rr]^{\delta_2} 
  \ar@<2.5ex>@{}[rr]|{(1)} &
\functF\,(\functM\beta)\,(\functM\beta)
  \ar[r]_{\delta}
  \ar[d]^(0.65){\contents_{\functF}}
  \ar[ddl]^(0.6){\functF\,(\oplus/)\,(\oplus/)} 
  \ar@{}[dr]|{(6)} &
\functM(\functF\,\beta\,\beta)
  \ar[dr]^{\functM f}
  \ar[d]_(0.35){\functM\contents_{\functF}} &
{}
  \ar@{}[dl]|(0.7){(2)} \\
{}
  \ar@{}[dr]|(0.6){(5)} &
[\functM\beta] 
  \ar[r]_{\delta}
  \ar[d]^{\functList(\oplus/)} 
  \ar@{}[drr]|{(7)} &
\functM[\beta]
  \ar[r]_{\functM(\foldr\,(\otimes)\,b)} &
\functM\beta
  \ar[d]^{\oplus/} \\
\functF\,\beta\,\beta
  \ar[r]_{\contents_{\functF}} 
  \ar@/_3pc/[rrr]_{f} 
  \ar@<-3ex>@{}[rrr]|{(3)} &
[\beta]
  \ar[rr]_{\foldr\,(\otimes)\,b} &
&
\beta
} \]
The perimeter is just the commuting diagram given in 
Section~\ref{sec:horner}---the diagram we have to justify. Face~(1) is the
definition of $\delta_2$ in terms of $\delta$. Faces~(2) and (3) are
the expansion of $f$ as generic folding of an
$\functF$-structure. Face~(4) follows from $\oplus/$ being an
$\functM$-algebra, and hence being a left-inverse of
$\return$. Face~(5) is an instance of the naturality property of
$\contents_{\functF} : \functF\triangle \ntto
\functList$. Face~(6) is the property that $\delta$ respects the
contents-preserving transformation $\contents_{\functF}$. 
\label{use:diagram-chase}
Therefore, the
whole diagram commutes if Face~(7) does---so let's focus on Face~(7):
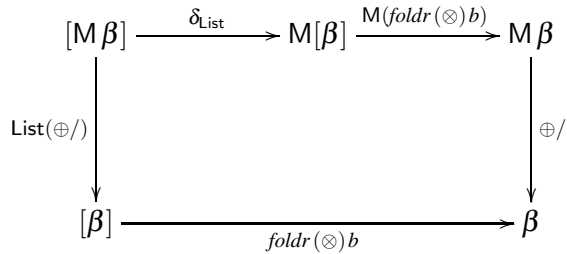
\[ \xymatrix@=0.75in{
[\functM\,\beta] \ar[r]^{\delta_{\functList}} \ar[d]_{\functList(\oplus/)} &
\functM[\beta] \ar[r]^{\functM(\foldr\,(\otimes)\,b)} &
\functM\,\beta \ar[d]^{\oplus/} \\
[\beta] \ar[rr]_{\foldr\,(\otimes)\,b} && 
\beta
} \]
Demonstrating that this diagram commutes is not too difficult, because
both sides turn out to be list folds. 
Around the left and bottom edges, we have
a fold $\foldr\,(\otimes)\,b$ after a map $\functList\,(\oplus/)$,
which automatically fuses to $\foldr\,(\odot)\,b$, where $\odot$ is defined by
$x \odot a = (\oplus/x) \otimes a$,
or, pointlessly, 
$(\odot) = (\otimes) \cdot (\oplus/) \times \id$.
Around the top and right edges we have the composition $\oplus/ \cdot
\functM(\foldr\,(\otimes)\,b) \cdot \delta_{\functList}$. If we can write
$\delta_{\functList}$ as an instance of $\foldr$, we can then use the
fusion law for $\foldr$ 
to prove that this composition equals $\foldr\,(\odot)\,b$
(see Exercise~\ref{ex:distr-fusion}). \label{use:distr-fusion}

In fact, there are various equivalent ways of writing
$\delta_{\functList}$ as an instance of $\foldr$. The definition given
by Conor McBride and Ross Paterson in their
original paper on
  idioms
\cite{McBride&Paterson2008:Applicative}
looked like the identity function, but with added idiomness:
\[ \begin{array}{@{}lcl}
\delta_{\functList}\,[\,] &=& \mathit{pure}\,[\,] \\
\delta_{\functList}\,(\mathit{mb} : \mathit{mbs}) &=& \mathit{pure}\,(:) \circledast \mathit{mb} \circledast \delta_{\functList}\,\mathit{mbs}
\end{array} \]
In the special case that the idiom is a monad, it can be written in terms of $\mathit{liftM}_0$ (aka $\return$) and $\mathit{liftM}_2$:
\[ \begin{array}{@{}lcl}
\delta_{\functList}\,[\,] &=& \mathit{liftM}_0\,[\,] \\
\delta_{\functList}\,(\mathit{mb} : \mathit{mbs}) &=& \mathit{liftM}_2\,(:)\,\mathit{mb}\,(\delta_{\functList}\,\mathit{mbs})
\end{array} \]
But we'll use a third equivalent definition:
\[ \begin{array}{@{}lcl}
\delta_{\functList}\,[\,] &=& \return\,[\,] \\
\delta_{\functList}\,(\mathit{mb} : \mathit{mbs}) &=& \functM(:)\,(\mathit{cp}\,(\mathit{mb}, \delta_{\functList}\,\mathit{mbs}))
\end{array} \]
where
\[ \begin{array}{@{}lcl}
\mathit{cp} &::& \functM\,\alpha \times \functM\,\beta \to \functM(\alpha\times\beta) \\
\mathit{cp}\,(x,y) &=& \join\,(\functM\,(\lambda a \mathbin{.} \functM\,(a,)\,y)\,x)
\end{array} \]
That is,
\[ \delta_{\functList} = \foldr\,(\functM(:)\cdot\mathit{cp})\,(\return\,[\,]) \]
In the use of fold fusion in demonstrating distributivity for lists
(Exercise~\ref{ex:distr-fusion}), we are naturally lead to a
distributivity condition
\[\oplus/ \cdot \functM(\otimes) \cdot \mathit{cp} = (\otimes) \cdot (\oplus/)\times(\oplus/) \]
for $\mathit{cp}$. This in turn follows from corresponding
distributivity properties for collections 
\label{use:distr-cp} (see Exercise~\ref{ex:distr-cp}),
\[ \begin{array}{@{}lcl}
\oplus/ \cdot \functM(a\otimes) &=& (a\otimes) \cdot \oplus/ \\
\oplus/ \cdot \functM(\otimes b) &=& (\otimes b) \cdot \oplus/ \\
\end{array} \]
which can finally be discharged by induction over the size of the (finite!)
collections (see Exercise~\ref{ex:distr-coll}). \label{use:distr-coll}

\section{Conclusion} \label{sec:relations}

As the title of their paper \cite{Bird*96:Generic} suggests, Bird \&
co carried out their development using the relational approach set out
in the \textit{Algebra of Programming} book
\cite{Bird&deMoor96:Algebra}; for example, their version of
$\mathit{prune}$ is a relation between data structures and their
prunings, rather than being a function that takes a structure and
returns the collection of all its prunings. There's a well-known
isomorphism between relations and set-valued functions, so their
relational approach roughly looks equivalent to the monadic one taken
here.

I've known their paper well for over a decade (I made essential use of the
"labelled variant" construction in my own papers on generic downwards
accumulations
\cite{Gibbons98:Polytypic,Gibbons2000:Generic}), 
but I've only just noticed that although they discuss
the maximum segment sum problem, they don't discuss problems based on
other semirings, such as the obvious one of integers with addition and
multiplication---which is, after all, the origin of Horner's Rule. Why
not? It turns out that the relational approach doesn't work in that
case!

There's a hidden condition in the calculation, which relates back to
our earlier comment about which collection monad---finite sets, finite
bags, lists, etc---to use. When $\functM$ is the set monad,
distribution over choice ($\oplus / (x \mplus y) = (\oplus/x) \oplus
(\oplus/y)$)---and consequently the condition ${\oplus/} \cdot \opt\,b
= (b\oplus) \cdot {\oplus/}$ that we used in proving Horner's
Rule---requires $\oplus$ to be idempotent, because $\mplus$
itself is idempotent; but addition is not idempotent. For exactly this
reason, the distributivity property does not in fact hold for addition with
the set monad. But everything does work out with the bag monad, for
which $\mplus$ is not idempotent. The bag monad models a
flavour of nondeterminism in which multiplicity of results
matters---as it does for the sum-of-products instance of the problem,
when two copies of the same segment should be treated differently from
just one copy. Similarly, if the order of results matters---if, for
example, we were looking for the "first" solution---then we would have
to use the list monad rather than bags or sets.
The moral of the story is that the relational approach is
\emph{programming with just one monad}, namely the set monad; if that
monad doesn't capture your effects faithfully, you're stuck.

(On the other hand, there are aspects of the problem that work much
better relationally than they do functionally. We have carefully used maximum only for a linear
order, namely the usual ordering of the integers. A non-antisymmetric order is
more awkward monadically, because there need not be a unique maximal
value. For example, it is not so easy to compute "the" segment with
maximal sum, because there may be several such. We could refine the 
ordering by sum on segments to make it once more a partial order,
perhaps breaking ties lexically; but we have to take care
to preserve the right distributivity properties.
Relationally, however, finding the maximal elements of a finite
collection under a partial order works out perfectly
straightforwardly.  We can try the same trick of turning the relation
"maximal under a partial order" into the collection-valued function
"all maxima under a partial order", but the equivalent
trick on the ordering itself---turning the relation "$<$" into the
collection-valued function "all values less than this one"---runs into
problems by taking us outside the world of \emph{finite}
nondeterminism.)

{\catcode`\_=\active \def_{\_}
\bibliographystyle{eptcs}
\bibliography{mssm}
}

\section{Appendix: Notation} \label{sec:notation}

For the benefit of those not fluent in Haskell and the Algebra of
Programming approach, this appendix presents some basic notations. For
a more thorough introduction, see the books by Richard Bird
\cite{Bird98:Introduction,Bird&deMoor96:Algebra} and my lecture notes
on "origami programming" \cite{Gibbons2003:Origami}.

\begin{description}

\item[Types:] Our programs are typed; the statement "$x :: \alpha$"
  declares that variable or expression $x$ has type~$\alpha$. We use
  product types $\alpha \times \beta$
  (with morphism $\fork :: (\alpha\to\beta)\times(\alpha\to\gamma) \to (\alpha\to\beta\times\gamma)$),
  sum types $\alpha + \beta$, and
  function types $\alpha \to \beta$. We assume throughout that types
  represent sets, and functions are total. 

\item[Functions:] Function application is usually denoted by
  juxtaposition, "$f\,x$", and is left-associative and
  tightest-binding. Function composition is backwards, so
  $(f \cdot g)\,x = f\,(g\,x)$. 

\item[Operators:] It is often convenient to write binary operators in
  infix notation; this makes many algebraic equations more
  perspicuous. We use sections $(a\oplus)$ and $(\oplus b)$ for
  partially applied binary operators, so that $(a\oplus)\,b = a \oplus
  b = (\oplus b)\,a$. In contrast to Haskell, we consider binary
  operators uncurried; for example, $(+) :: \Z \times \Z \to \Z$.

\item[Lists:] We use the Haskell syntax "$[\alpha]$" for a list type,
  "$[\,]$" for the empty list, "$a:x$" for cons, "$\append$" for append,
  and "$[1,2,3]$" for a
  list constant. The fold $\foldr :: (\alpha \times \beta \to \beta)
  \to \beta \to [\alpha] \to \beta$ is ubiquitous; it has the
  universal property
  \[ h = \foldr\,f\,e \quad\Leftrightarrow\quad
     h\,[\,]=e \;\land\; h \cdot (:) = f \cdot \id\times h \]
  and as a special case of this, the fusion law
  \[
  h \cdot \foldr\,f\,e = \foldr\,f'\,e' \quad\Leftarrow\quad
  h\,e=e' \;\land\; h \cdot f = f' \cdot \id\times h
  \]
  The function $\map :: (\alpha\to\beta) \to [\alpha] \to [\beta]$
  is an instance, via
  $\map\,f = \foldr\,((:) \cdot f \times \id)\,[\,]$.
  So is $\scanr$, which computes the fold of every tail of a list:
  \[ \begin{array}{@{}lcl}
  \scanr &::& (\alpha \times \beta \to \beta) \to \beta \to [\alpha] \to [\beta] \\
  \scanr\,f\,e &=& \foldr\,h\,[e]
  \quad \where\; h\,a\,(b:x) = f\,a\,b : (b:x)
  \end{array} \]
  We also use the variant $\foldr_1\,f\,(x \append [a]) = \foldr\,f\,a\,x$
  on non-empty lists.

\item[Functors:] Datatypes are modelled as functors, which are
  operations on both types and functions; so for $\functF$ a functor,
  $\functF\,\alpha$ is a type whenever $\alpha$ is, and if $f ::
  \alpha\to\beta$ then $\functF\,f :: \functF\,\alpha \to
  \functF\,\beta$. Moreover, $\functF$ respects the compositional
  structure of functions, preserving identity ($\functF\,\id_\alpha =
  \id_{\functF\alpha}$) and composition ($\functF\,(f \cdot g) =
  \functF\,f \cdot \functF\,g$). For example, $\functList$ is a
  functor, with $\functList\,\alpha = [\alpha]$ and $\functList\,f =
  \map\,f$. We generalize this also to bifunctors, which are binary
  operators functorial in each argument; for example, we will see the
  bifunctor $\functL\,\alpha\,\beta = 1 + \alpha \times \beta$ below,
  as the "shape functor" for lists.

\item[Naturality:] Polymorphic functions are modelled as natural
  transformations between functors.  A natural transformation $\phi :
  \functF \ntto \functG$ is a family of functions $\phi_\alpha ::
  \functF\,\alpha \to \functG\,\alpha$, one for each $\alpha$, coherent in
  the sense of being related by the naturality condition $\functG\,h \cdot
  \phi_\alpha = \phi_\beta \cdot \functF\,h$ whenever $h :: \alpha \to
  \beta$.

\item[Datatype-genericity:] Datatype-generic programming is expressed
  in terms of parametrization by a functor. In particular, for a large
  class of bifunctors $\functF$ (including all those built from
  constants and the identity using sums and products---the polynomial
  bifunctors), we can form a kind of least fixed point
  $\functT\,\alpha = \mu(\functF\,\alpha)$ of $\functF$ in its second
  argument, giving an inductive datatype.
  It is a "fixed point" in the sense that $\functT\,\alpha
  \simeq \functF\,\alpha\,(\functT\,\alpha)$; so $\functList\,\alpha =
  \mu(\functL\,\alpha)$, where $\functL$ is the shape functor for
  lists defined above.  
  We sometimes use Haskell-style datatype definitions, which conveniently 
  name the constructors too:
  \[ \data\;\functList\,\alpha = \mathit{Nil} \mid \mathit{Cons}\,(\alpha, \functList\,\alpha) \]

\item[Algebras:] An $\functF$-algebra is a pair $(\alpha,f)$ such that
  $f :: \functF\,\alpha \to \alpha$. A homomorphism between
  $\functF$-algebras $(\alpha,f)$ and $(\beta,g)$ is a function $h ::
  \alpha \to \beta$ such that $h \cdot f = g \cdot \functF\,h$.
  One half of the isomorphism by which an inductive datatype is a 
  fixed point is given by the
  constructor $\inn_\functF :: \functF\,\alpha\,(\functT\,\alpha) \to
  \functT\,\alpha$, through which $(\functT\,\alpha, \inn_\functF)$
  forms an $(\functF\,\alpha)$-algebra. The datatype is the "least" fixed point
  in the sense that there is a unique homomorphism to any other
  $(\functF\,\alpha)$-algebra $(\beta,f)$; we say that
  $(\functT\,\alpha, \inn_\functF)$ is the initial
  $(\functF\,\alpha)$-algebra. We write $\fold_\functF\,f$ for that
  unique homomorphism; its uniqueness is captured in the universal property
  \[ h = \fold_\functF\,f \quad\Leftrightarrow\quad h \cdot \inn_\functF = f \cdot \functF\,h \]

\item[Monads:] A monad $\functM$ is a functor with two additional
  natural transformations, a multiplication $\join : \functM\functM
  \ntto \functM$ and a unit $\return : \functId \ntto \functM$ (where $\functId$
  is the identity functor), that satisfy three laws:
  \[ \begin{array}{@{}lcl}
  \join \cdot \return &=& \id \\
  \join \cdot \functM\,\return &=& \id \\
  \join \cdot \functM\,\join &=& \join \cdot \join
  \end{array} \]
  Collection types such as finite lists, bags, and sets form monads; in
  each case, $\return$ yields a singleton collection, and $\join$ unions
  a collection of collections into a collection. Another monad we will
  use is Haskell's "maybe" datatype and associated morphism
  \[ \begin{array}{@{}lcl}
  \rlap{$\data\;\functMaybe\,\alpha = \Nothing \mid \Just\,\alpha$} \smallskip \\
  \maybe\,e\,f\,\Nothing &=& e \\
  \maybe\,e\,f\,(\Just\,a) &=& f\,a
  \end{array} \]
  for which $\return = \Just$ and $\join = \maybe\,\Nothing\,\id$.
  An algebra for a monad $\functM$ is an $\functM$-algebra $(\alpha,f)$ for
  $\functM$ as a functor, satisfying the extra conditions
  \[ \begin{array}{@{}lcl}
  f \cdot \return &=& \id \\
  f \cdot \join &=& f \cdot \functM\,f
  \end{array} \]

\item[Idioms:] An idiom $\functM$ is a functor with two additional
  natural transformations, whose components are
  $\mathit{pure}_\alpha :: \alpha \to \functM\,\alpha$
  and ${\circledast}_{\alpha,\beta} ::
  \functM\,(\alpha\to\beta)\times\functM\,\alpha\to\functM\,\beta$,
  satisfying four laws:
  \[ \begin{array}{@{}lcl}
  \mathit{pure}\,\id \circledast u &=& u \\
  \mathit{pure}\,(\cdot) \circledast u \circledast v \circledast w &=& u \circledast (v \circledast w) \\
  \mathit{pure}\,f \circledast \mathit{pure}\,a &=&  \mathit{pure}\,(f a) \\
  u \circledast \mathit{pure}\,a &=&  \mathit{pure}\,(\lambda f \mathbin{.} f\,a) \circledast u
  \end{array} \]
  Any monad induces an idiom; so does any constant functor
  $\functK_\alpha$, provided that there is a monoidal structure on $\alpha$.

\end{description}

\section{Appendix: Exercises} \label{sec:exercises}

\ifanswers\raggedbottom\fi

\begin{enumerate}

\item (See page~\pageref{use:mss-lists}.) \label{ex:mss-lists}
Calculate that
\[ 
\mathit{mss} = \maximum \cdot \map\,(\maximum \cdot \map\,\sum \cdot \inits) \cdot \tails
\]
just using the definitions of $\mathit{mss},\inits,\tails$,
together with (i)~distributivity of $\map$ over
function composition, (ii)~naturality of $\concat$, that is, $\map\,f
\cdot \concat = \concat \cdot \map\,(\map\,f)$, and (iii)~that
$\maximum$ is a list homomorphism, that is, $\maximum \cdot \concat =
\maximum \cdot \map\,\maximum$.
\ifanswers \begin{answer}
\[ \begin{array}{@{}ll}
  & \mathit{mss} \\
= & \qquad \{\; \mbox{definition of \(\mathit{mss}\)} \;\} \\
  & \maximum \cdot \map\,\sum \cdot \segs \\
= & \qquad \{\; \mbox{definition of \(\segs\)} \;\} \\
  & \maximum \cdot \map\,\sum \cdot \concat \cdot \map\,\inits \cdot \tails \\
= & \qquad \{\; \mbox{naturality: \(\map\,f \cdot \concat = \concat \cdot \map\,(\map\,f)\)} \;\} \\
  & \maximum \cdot \concat \cdot \map\,(\map\,\sum) \cdot \map\,\inits \cdot \tails \\
= & \qquad \{\; \mbox{homomorphisms: \(\maximum \cdot \concat = \maximum \cdot \map\,\maximum\)} \;\} \\
  & \maximum \cdot \map\,\maximum \cdot \map\,(\map\,\sum) \cdot \map\,\inits \cdot \tails \\
= & \qquad \{\; \mbox{functors} \;\} \\
  & \maximum \cdot \map\,(\maximum \cdot \map\,\sum \cdot \inits) \cdot \tails
\end{array} \]
\end{answer} \fi

\item (See page~\pageref{use:horner}.) \label{ex:horner}
Use the sum-of-products version of Horner's Rule to prove the more familiar 
polynomial version.
\ifanswers \begin{answer}
In the case that all the $a_i$ are non-zero, we have
\[ \displaystyle
\sigmasum_{i=0}^{n-1} a_i x^i = a_0 \sigmasum_{i=0}^{n-1} \piprod_{j=0}^{i-1} a_{j+1}x/a_j
\]
In general, we have to skip the terms $a_ix^i$ when $a_i=0$, but that works out fine; for example,
when $a_1=0$ but $a_0,a_2,a_3 \ne 0$, we have
\[ \displaystyle
a_0 + a_1x + a_2x^2 + a_3x^3 = a_0 (1 + u_0 + u_0u_1)
\]
where $u_0 = a_2x^2/a_0$ and $u_1 = a_3x/a_2$.
\end{answer} \fi

\item (See page~\pageref{use:simulate}.) \label{ex:simulate}
Hand-simulate the execution of the linear-time algorithm for $\mathit{mss}$
\[ \mathit{mss} = \foldr\,(\oplus)\,e \quad
\where\; e = 0 \mathbin{;} u \oplus z = e \max (u + z) \]
on the list $[4, -5, 6, -3, 2, 0, -4, 5, -6, 5]$. Do you understand how it works?
\ifanswers \begin{answer}
\[ \begin{array}{@{}ll}
  & \foldr\,(\oplus)\,e\,[4, -5, 6, -3, 2, 0, -4, 5, -6, 5] \\
= & 4 \oplus (-5 \oplus (6 \oplus (-3 \oplus (2 \oplus (0 \oplus (-4 \oplus (5 \oplus (-6 \oplus (5 \oplus 0))))))))) \\
= & 4 \oplus (-5 \oplus (6 \oplus (-3 \oplus (2 \oplus (0 \oplus (-4 \oplus (5 \oplus (-6 \oplus 5)))))))) \\
= & 4 \oplus (-5 \oplus (6 \oplus (-3 \oplus (2 \oplus (0 \oplus (-4 \oplus (5 \oplus 0))))))) \\
= & 4 \oplus (-5 \oplus (6 \oplus (-3 \oplus (2 \oplus (0 \oplus (-4 \oplus 5)))))) \\
= & 4 \oplus (-5 \oplus (6 \oplus (-3 \oplus (2 \oplus (0 \oplus 1))))) \\
= & 4 \oplus (-5 \oplus (6 \oplus (-3 \oplus (2 \oplus 1)))) \\
= & 4 \oplus (-5 \oplus (6 \oplus (-3 \oplus 3))) \\
= & 4 \oplus (-5 \oplus (6 \oplus 0)) \\
= & 4 \oplus (-5 \oplus 6) \\
= & 4 \oplus 1 \\
= & 5
\end{array} \]
\end{answer} \fi

\item (See page~\pageref{use:semirings}.) \label{ex:semirings}
Apart from $({+},{\times})$ and $({\max},{+})$, what other semirings do you know, and what variations on the "maximum segment sum" problem do they suggest?
\ifanswers \begin{answer}
Here are a few more semirings:
\begin{itemize}
\item $({\sqcap},{+})$ works just as well as $({\max},{+})$
\item booleans with $({\lor},{\land})$
\item sets with $({\cup},{\cap})$
\item square matrices of integers, with addition and multiplication
\item indeed, square matrices with elements drawn from any semiring (such as the booleans)
\item Kleene algebras---eg languages (sets of strings) with union and concatenation
\end{itemize}
However, semirings that are also lattices, such as $({\land},{\lor})$ and $({\cap},{\cup})$, are not very interesting in this context.
\end{answer} \fi

\item (See page~\pageref{use:labelled}.) \label{ex:labelled}
Verify that the labelled variant of the usual datatype of lists
(namely, $\functList\,\alpha = \mu(\functF\,\alpha)$ where shape functor 
$\functF$ is given by
$\functF\,\alpha\,\beta = 1 + \alpha \times \beta$) is a datatype of nonempty 
lists. What is the labelled variant of externally-labelled binary trees, 
whose shape functor is $\functF\,\alpha\,\beta = \alpha + \beta \times \beta$?
That of internally-labelled binary trees, whose shape functor is 
$\functF\,\alpha\,\beta = 1 + \alpha \times \beta \times \beta$?
And homogeneous binary trees, whose shape functor is 
$\functF\,\alpha\,\beta = \alpha + \alpha \times \beta \times \beta$?
\ifanswers \begin{answer}
For lists, the construction in the text gives labelled variant $\functL\,\alpha = \mu(\functG\,\alpha)$ where
$\functG\,\alpha\,\beta = \alpha \times \functF\,1\,\beta = \alpha \times (1 + 1 \times \beta) \simeq \alpha + \alpha \times \beta$,
which does indeed give a datatype of nonempty lists. For each of the three kinds of binary tree mentioned, we end up with
$\functG\,\alpha\,\beta = \alpha + \alpha \times \beta \times \beta$,
the shape of homogeneous binary trees.
\end{answer} \fi

\item (See page~\pageref{use:para-unfold}.) \label{ex:para-unfold}
If you're familiar with paramorphisms and with anamorphisms (unfolds),
write $\mathit{subterms}_\functF$ and $\mathit{scan}_\functF$ as instances of these.
\ifanswers \begin{answer}
Paramorphisms are like catamorphisms (folds), except that at each
step we have access to the original subterms as well as the result
of the recursive call on those subterms. So whereas $\fold_\functF$
takes a body of type $\functF\,\alpha\,\beta\to\beta$, the
paramorphism $\mathit{para}_\functF$ takes a body of type
$\functF\,\alpha\,(\beta\times\mu(\functF\,\alpha))\to\beta$. This
is convenient for $\mathit{subterms}$, because it saves us from having to
reconstruct those original subterms using $\mathit{root}$:
\[
\mathit{subterms}_\functF = \mathit{para}_\functF(\inn_{\functG} \cdot \fork(\inn_{\functF} \cdot \functF\,\id\,\snd, \functF\,!\,\fst))
\]
Sadly, it isn't so convenient for $\mathit{scan}$, because the
original subterms that we have carefully saved still have to be
folded:
\[
\mathit{scan}_\functF\,f =
\fold_{\functF}(\inn_{\functG} \cdot \fork(f \cdot \functF\,\id\,(\fold_\functF\,f\cdot\snd), \functF\,!\,\fst))
\]
Unfolds are the dual of folds; whereas the fold has type
$(\functF\,\alpha\,\beta\to\beta) \to (\mu(\functF\,\alpha)\to\beta)$,
the unfold has type $(\beta\to\functF\,\alpha\,\beta) \to
(\beta\to\mu(\functF\,\alpha))$, reversing two of the arrows.
(Technically, the unfold generates an element of the \emph{greatest}
fixpoint type $\nu(\functF\,\alpha)$ instead of the least fixpoint
$\mu(\functF\,\alpha)$; but in Haskell, based on complete partial
orders, these coincide.)  Again, this works quite nicely for
$\mathit{subterms}$: the root is a copy of the whole structure, and
the children are generated recursively from children of the input.
\[
\mathit{subterms}_\functF =
\mathit{unfold}_\functG(\fork(\id, \functF\,!\,\id \cdot \out))
\]
However, it is rather inefficient to define $\mathit{scan}$ as an
unfold, because there is no sharing of common subexpressions:
\[
\mathit{scan}_\functF\,f =
\mathit{unfold}_\functG(\fork(\fold_\functF\,f, \functF\,!\,\id \cdot \out))
\]
\end{answer} \fi

\item (See page~\pageref{use:prune}.) \label{ex:prune}
Hand-simulate the execution of $\mathit{prune}$ in the finite bag monad 
on a small homogeneous binary tree, such as the term
$\mathit{Fork}\,(1, \mathit{Leaf}\,2, \mathit{Fork}\,(3, \mathit{Leaf}\,1, \mathit{Leaf}\,4))$
of type
\[
\data\;\functTree\,\alpha = \mathit{Leaf}\,\alpha \mid \mathit{Fork}\,(\alpha,\functTree\,\alpha,\functTree\,\alpha)
\]
What happens on externally-labelled binary trees? Internally-labelled?
How does the result differ if you let $\functM$ be sets rather than bags?
\ifanswers \begin{answer}
Let's define concrete syntax for the datatype of prunable homogeneous trees, which have one more constructor:
\[
\data\;\functU\,\alpha = E \mid L\,\alpha \mid F\,(\alpha,\functU\,\alpha,\functU\,\alpha)
\]
Writing $\langle\ldots\rangle$ for a bag, 
and making use of a "bag comprehension" notation $\langle e \mid a \leftarrow x \rangle$,
we can unpack the definition of $\mathit{prune}$ to 
reveal the following recursive characterization:
\[ \begin{array}{@{}lcl}
\mathit{prune}\,(\mathit{Leaf}\,a) &=& \langle E, L\,a \rangle \\
\mathit{prune}\,(\mathit{Fork}\,(a,t,u)) &=& \langle E \rangle \uplus 
\langle F\,(a,t',u') \mid t' \leftarrow \mathit{prune}\,t, u' \leftarrow \mathit{prune}\,u \rangle
\end{array} \]
In particular, the three-element subtree rooted at $3$ has five prunings,
and the whole five-element tree has eleven prunings:
\[ \begin{array}{@{}ll} \langle &
E,
F\,(1, E, E),
F\,(1, E, F\,(3, E, E)),
F\,(1, E, F\,(3, E, L\,4)),
F\,(1, E, F\,(3, L\,1, E)), \\ &
F\,(1, E, F\,(3, L\,1, L\,4)),
F\,(1, L\,2, E),
F\,(1, L\,2, F\,(3, E, E)), \\ &
F\,(1, L\,2, F\,(3, E, L\,4)),
F\,(1, L\,2, F\,(3, L\,1, E)),
F\,(1, L\,2, F\,(3, L\,1, L\,4))
\quad \rangle \end{array} \]
\end{answer} \fi

\item (See page~\pageref{use:distr-sum-max}.) \label{ex:distr-sum-max}
Pick a shape functor $\functF$ and a collection monad $\functM$;
give suitable definitions of 
$f :: \functF\,\N\,\N \to \N$ to sum all
naturals in an $\functF$-structure and $k :: \functM\,\N \to \N$
to find the maximum of a collection of naturals;
and verify that the rectangle in Section~\ref{sec:horner} commutes.
\ifanswers \begin{answer}
Here's how things work out for lists, using the bag monad, and the 
concrete syntax
\[
\data\;\functF\,\alpha\,\beta = N \mid C\,\alpha\,\beta
\]
for the shape functor $\functF\,\alpha\,\beta = 1 + \alpha \times
\beta$. We define
\[ \begin{array}{@{}lcl}
f\,N &=& 0 \\
f\,(C\,n\,m) &=& n+m
\end{array} \]
to sum an $\functF$-structure of naturals, and let $k = \mathit{max}$,
the obvious function to compute the maximum of a bag of naturals
(returning $0$, the minimum natural, for the empty bag).  
Using the bag comprehension notation from Exercise~\ref{ex:prune},
the distributor $\delta_2$ is given by
\[ \begin{array}{@{}lcl}
\delta_2\,N &=& \mzero \\
\delta_2\,(C\,n\,x) &=& \langle C\,n\,m \mid m \leftarrow x \rangle
\end{array} \]
We have to show that
\[
\mathit{max} \cdot \functM\,f \cdot \delta_2 = f \cdot \functF\,\id\,\mathit{max}
\]
which we do by case analysis on the $\functF$-structured argument---for the nil case,
\[ \begin{array}{@{}ll}
  & \mathit{max}\,(\functM\,f\,(\delta_2\,N)) \\
= & \qquad \{\; \mbox{definition of \(\delta_2\)} \;\} \\
  & \mathit{max}\,(\functM\,f\,\mzero) \\
= & \qquad \{\; \mbox{functors} \;\} \\
  & \mathit{max}\,\mzero \\
= & \qquad \{\; \mbox{definition of \(\mathit{max}\)} \;\} \\
  & 0 \\
= & \qquad \{\; \mbox{definition of \(f\)} \;\} \\
  & f\,N \\
= & \qquad \{\; \mbox{functors} \;\} \\
  & f\,(\functF\,\id\,\mathit{max}\,N)
\end{array} \]
and for the cons case,
\[ \begin{array}{@{}ll}
  & \mathit{max}\,(\functM\,f\,(\delta_2\,(C\,n\,x))) \\
= & \qquad \{\; \mbox{definition of \(\delta_2\)} \;\} \\
  & \mathit{max}\,(\functM\,f\,\langle C\,n\,m \mid m \leftarrow x \rangle) \\
= & \qquad \{\; \mbox{functors} \;\} \\
  & \mathit{max}\,\langle f\,(C\,n\,m) \mid m \leftarrow x \rangle \\
= & \qquad \{\; \mbox{definition of \(f\)} \;\} \\
  & \mathit{max}\,\langle n+m \mid m \leftarrow x \rangle \\
= & \qquad \{\; \mbox{addition distributes over maximum} \;\} \\
  & n + \mathit{max}\,x \\
= & \qquad \{\; \mbox{definition of \(f\)} \;\} \\
  & f\,(C\,n\,(\mathit{max}\,x))\\
= & \qquad \{\; \mbox{functors} \;\} \\
  & f\,(\functF\,\id\,\mathit{max}\,(C\,n\,x))
\end{array} \]
\end{answer} \fi

\item (See page~\pageref{use:binary}.) \label{ex:binary}
Given an $\functM$-algebra $k$, show that $k$ distributes over 
$\mplus$---there exists a binary operator $\oplus$ such that 
$k\,(x \mplus y) = k\,x \oplus k\,y$.
(Hint: define $a \oplus b = k\,(\return\,a \mplus \return\,b)$.)
\ifanswers \begin{answer}
Defining $\oplus$ following the hint, we have:
\[ \begin{array}{@{}ll}
  & k\,x \oplus k\,y \\
= & \qquad \{\; \mbox{definition of \(\oplus\)} \;\} \\
  & k\,(\return\,(k\,x) \mplus \return\,(k\,y)) \\
= & \qquad \{\; \mbox{naturality of \(\return\)} \;\} \\
  & k\,(\functM\,k\,(\return\,x) \mplus \functM\,k\,(\return\,y)) \\
= & \qquad \{\; \mbox{naturality of \(\mplus\)} \;\} \\
  & k\,(\functM\,k\,(\return\,x \mplus \return\,y)) \\
= & \qquad \{\; \mbox{\(k\) a monad algebra} \;\} \\
  & k\,(\join\,(\return\,x \mplus \return\,y)) \\
= & \qquad \{\; \mbox{\(\join\) distributes over \(\mplus\)} \;\} \\
  & k\,(\join\,(\return\,x) \mplus \join\,(\return\,y)) \\
= & \qquad \{\; \mbox{monads} \;\} \\
  & k\,(x \mplus y)
\end{array} \]
\end{answer} \fi

\item (See page~\pageref{use:properties}.) \label{ex:properties}
Given an $\functM$-algebra $\oplus/$,
show that $\oplus$ is associative if $\mplus$ is; similarly for 
commutativity and idempotence. Show also that if $\mplus$ has a unit $\mzero$,
then $\oplus$ also has a unit, which must equal $\oplus/\mzero$.
(Hint: show first that $\oplus/$ is surjective.)
\ifanswers \begin{answer}
Given the distributivity of $\oplus/$ over $\mplus$ from
Exercise~\ref{ex:binary}, it is straightforward to show that $\oplus$
is associative on the range of $\oplus/$, provided that $\mplus$ is
itself associative:
\[ 
  \oplus/x \oplus (\oplus/y \oplus \oplus/z)
= \oplus/\,(x \mplus (y \mplus z))
= \oplus/\,((x \mplus y) \mplus z)
= (\oplus/x \oplus \oplus/y) \oplus \oplus/z
\]
Moreover, because $\oplus/$ is an $\functM$-algebra, we have $\oplus/
\cdot \return = \id$ and hence $\oplus/$ is surjective---so $\oplus$
is in fact associative everywhere. (In other words, just replay the
above calculation with $x = \return\,a$ etc.) The same argument works
for commutativity, idempotence, and unit laws.
\end{answer} \fi

\item (See page~\pageref{use:dgp-horner}.) 
Using the universal property of $\fold_\functF$, prove the fold fusion law
\[ h \cdot \fold_\functF\,f = \fold_\functF\,g \quad\Leftarrow\quad h \cdot f = g \cdot \functF\,h \]
Use this to prove the special case of fold--map fusion
\[ \fold_\functF\,f \cdot \functT\,g = \fold_\functF\,(f \cdot \functF\,g\,\id) \]
where $\functT\,\alpha = \mu(\functF\,\alpha)$.
\ifanswers \begin{answer}
For the fold fusion law, we have:
\[ \begin{array}{@{}ll}
  & h \cdot \fold_\functF\,f = \fold_\functF\,g \\
\Leftrightarrow 
  & \qquad \{\; \mbox{universal property} \;\} \\
  & h \cdot \fold_\functF\,f \cdot \inn_\functF = g \cdot \functF\,(h \cdot \fold_\functF\,f) \\
\Leftrightarrow 
  & \qquad \{\; \mbox{functors} \;\} \\
  & h \cdot \fold_\functF\,f \cdot \inn_\functF = g \cdot \functF\,h \cdot \functF\,(\fold_\functF\,f) \\
\Leftrightarrow 
  & \qquad \{\; \mbox{evaluation rule} \;\} \\
  & h \cdot f \cdot \functF\,(\fold_\functF\,f) = g \cdot \functF\,h \cdot \functF\,(\fold_\functF\,f) \\
\Leftarrow 
  & \qquad \{\; \mbox{Leibniz} \;\} \\
  & h \cdot f = g \cdot \functF\,h
\end{array} \]
The map operation $\functT\,g$ is an instance of fold: with $\functF$
now a bifunctor, we have
\[ \functT\,g = \fold_{\functF\alpha}\,(\inn_{\functF\alpha} \cdot \functF\,g\,\id) \]
Investigating the fusion condition of the body with $\fold\,f$:
\[ \begin{array}{@{}ll}
  & \fold_{\functF\alpha}\,f \cdot \inn_{\functF\alpha} \cdot \functF\,g\,\id \\
= & \qquad \{\; \mbox{evaluation rule} \;\} \\
  & f \cdot \functF\,\id\,(\fold_{\functF\alpha}\,f) \cdot \functF\,g\,\id \\
= & \qquad \{\; \mbox{functors} \;\} \\
  & f \cdot \functF\,g\,\id \cdot \functF\,\id\,(\fold_{\functF\alpha}\,f)
\end{array} \]
justifies the fold--map fusion law:
\[ \fold_{\functF\alpha}\,f \cdot \functT\,g = \fold_{\functF\alpha}\,(f \cdot \functF\,g\,\id) \]
\end{answer} \fi

\item (See page~\pageref{use:dgp-horner}.) \label{ex:dgp-horner}
Use fold fusion to calculate a characterization of
$\mathord{\oplus/} \cdot \functM(\fold_{\functH}(\maybe\,b\,f)) \cdot \mathit{prune}$
as a fold, assuming the distributivity property
${\oplus/} \cdot \functM\,f \cdot \delta_2 = f \cdot \functF\,\id\,(\oplus/)$.
\ifanswers \begin{answer}
\[ \begin{array}{@{}ll}
  & \mathord{\oplus/} \cdot \functM(\fold_{\functH}(g)) \cdot \mathit{prune} \cdot \inn_{\functF} \\
= & \qquad \{\; \mbox{evaluation for \(\mathit{prune}\)} \;\} \\
  & \mathord{\oplus/} \cdot \functM(\fold_{\functH}(g)) \cdot \functM\,\inn_{\functH} \cdot \opt\,\Nothing \cdot \functM\,\Just \cdot \delta_2 \cdot \functF\,\id\,\mathit{prune} \\
= & \qquad \{\; \mbox{functors; evaluation for \(\fold_{\functH}\)} \;\} \\
  & \mathord{\oplus/} \cdot \functM(g \cdot \functH\,\id\,(\fold_{\functH}(g))) \cdot \opt\,\Nothing \cdot \functM\,\Just \cdot \delta_2 \cdot \functF\,\id\,\mathit{prune} \\
= & \qquad \{\; \mbox{\(\functM\,h \cdot \opt\,a = \opt\,(h\,a) \cdot \functM\,h\)} \;\} \\
  & \mathord{\oplus/} \cdot \opt\,(g\,\Nothing) \cdot \functM(g \cdot \functH\,\id\,(\fold_{\functH}(g))) \cdot \functM\,\Just \cdot \delta_2 \cdot \functF\,\id\,\mathit{prune} \\
= & \qquad \{\; \mbox{functors; \(\Just :: \functF\,\alpha \ntto \functH\,\alpha\)} \;\} \\
  & \mathord{\oplus/} \cdot \opt\,(g\,\Nothing) \cdot \functM\,g \cdot \functM(\Just \cdot \functF\,\id\,(\fold_{\functH}(g))) \cdot \delta_2 \cdot \functF\,\id\,\mathit{prune} \\
= & \qquad \{\; \mbox{functors; \(\delta_2 :: (\functF\alpha)\functM \ntto \functM(\functF\alpha)\)} \;\} \\
  & \mathord{\oplus/} \cdot \opt\,(g\,\Nothing) \cdot \functM(g \cdot \Just) \cdot \delta_2 \cdot \functF\,\id\,(\functM(\fold_{\functH}(g))) \cdot \functF\,\id\,\mathit{prune} \\
= & \qquad \{\; \mbox{\(g = \maybe\,b\,f\)} \;\} \\
  & \mathord{\oplus/} \cdot \opt\,b \cdot \functM\,f \cdot \delta_2 \cdot \functF\,\id\,(\functM(\fold_{\functH}(g))) \cdot \functF\,\id\,\mathit{prune} \\
= & \qquad \{\; \mbox{\({\oplus/} \cdot \opt\,b = (b\oplus) \cdot {\oplus/}\)} \;\} \\
  & (b\oplus) \cdot {\oplus/} \cdot \functM\,f \cdot \delta_2 \cdot \functF\,\id\,(\functM(\fold_{\functH}(g))) \cdot \functF\,\id\,\mathit{prune} \\
= & \qquad \{\; \mbox{distributivity: \({\oplus/} \cdot \functM\,f \cdot \delta_2 = f \cdot \functF\,\id\,(\oplus/)\)} \;\} \\
  & (b\oplus) \cdot f \cdot \functF\,\id\,(\oplus/) \cdot \functF\,\id\,(\functM(\fold_{\functH}(g))) \cdot \functF\,\id\,\mathit{prune} \\
= & \qquad \{\; \mbox{functors} \;\} \\
  & (b\oplus) \cdot f \cdot \functF\,\id\,({\oplus/} \cdot \functM(\fold_{\functH}(g)) \cdot \mathit{prune})
\end{array} \]
Therefore,
\[ {\oplus/} \cdot \functM(\fold_{\functH}(\maybe\,b\,f)) \cdot \mathit{prune} = \fold_{\functF}((b\oplus) \cdot f)
\]
\end{answer} \fi

\item (See page~\pageref{use:dgp-mss}.) \label{ex:dgp-mss}
Show by calculation that
\[ {\oplus/} \cdot \functM(\fold_{\functH}(\maybe\,b\,f)) \cdot \segs =
\mathord{\oplus/} \cdot \mathit{contents}_{\functL} \cdot \mathit{scan}_{\functF}((b\oplus)\cdot f)
\]
\ifanswers \begin{answer}
\[ \begin{array}{@{}ll}
  & \mathord{\oplus/} \cdot \functM(\fold_{\functH}(\maybe\,b\,f)) \cdot \segs \\
= & \qquad \{\; \mbox{definition of \(\segs\)} \;\} \\
  & \mathord{\oplus/} \cdot \functM(\fold_{\functH}(\maybe\,b\,f)) \cdot \join \cdot \functM\,\mathit{prune} \cdot \mathit{contents}_{\functL} \cdot \mathit{subterms} \\
= & \qquad \{\; \mbox{naturality of \(\join :: \functM\functM \ntto \functM\); functors} \;\} \\
  & \mathord{\oplus/} \cdot \join \cdot \functM(\functM(\fold_{\functH}(\maybe\,b\,f)) \cdot\mathit{prune}) \cdot \mathit{contents}_{\functL} \cdot \mathit{subterms} \\
= & \qquad \{\; \mbox{\(\oplus/\) is an \(\functM\)-algebra; functors} \;\} \\
  & \mathord{\oplus/} \cdot \functM({\oplus/} \cdot \functM(\fold_{\functH}(\maybe\,b\,f)) \cdot\mathit{prune}) \cdot \mathit{contents}_{\functL} \cdot \mathit{subterms} \\
= & \qquad \{\; \mbox{naturality of \(\mathit{contents} :: \functL \ntto \functM\)} \;\} \\
  & \mathord{\oplus/} \cdot \mathit{contents}_{\functL} \cdot \functL({\oplus/} \cdot \functM(\fold_{\functH}(\maybe\,b\,f)) \cdot\mathit{prune}) \cdot \mathit{subterms} \\
= & \qquad \{\; \mbox{Horner's Rule} \;\} \\
  & \mathord{\oplus/} \cdot \mathit{contents}_{\functL} \cdot \functL(\fold_{\functF}((b\oplus)\cdot f)) \cdot \mathit{subterms} \\
= & \qquad \{\; \mbox{Scan Lemma} \;\} \\
  & \mathord{\oplus/} \cdot \mathit{contents}_{\functL} \cdot \mathit{scan}_{\functF}((b\oplus)\cdot f)
\end{array} \]
\end{answer} \fi

\item (See page~\pageref{use:diagram-chase}.) \label{ex:diagram-chase}
Convert the big commuting diagram in Section~\ref{sec:distr-lists} into an 
equational proof of the distributivity property
${\oplus/} \cdot \functM\,f \cdot \delta_2 = f \cdot \functF\,\id\,(\oplus/)$,
assuming the properties captured by each of the individual faces.
\ifanswers \begin{answer}
\[ \begin{array}{@{}ll}
  & {\oplus/} \cdot \functM\,f \cdot \delta_2 \\
= & \qquad \{\; \mbox{defining \(\delta_2\) in terms of \(\delta_\functF\)} \;\} \\
  & {\oplus/} \cdot \functM\,f \cdot \delta_\functF \cdot \functF\,\return\,\id \\
= & \qquad \{\; \mbox{defining \(f\) in terms of \(\contents\); functors} \;\} \\
  & {\oplus/} \cdot \functM\,(\foldr\,(\otimes)\,b) \cdot \functM\,\contents_{\functF} \cdot \delta_\functF \cdot \functF\,\return\,\id \\
= & \qquad \{\; \mbox{\(\delta\) respects \(\contents\)} \;\} \\
  & {\oplus/} \cdot \functM\,(\foldr\,(\otimes)\,b) \cdot \delta_\functList \cdot \contents_{\functF} \cdot \functF\,\return\,\id \\
= & \qquad \{\; \mbox{distributivity for lists (Exercise~\ref{ex:distr-fusion})} \;\} \\
  & \foldr\,(\otimes)\,b \cdot \functList\,(\oplus/) \cdot \contents_{\functF} \cdot \functF\,\return\,\id \\
= & \qquad \{\; \mbox{naturality of \(\contents\)} \;\} \\
  & \foldr\,(\otimes)\,b \cdot \contents_{\functF} \cdot \functF\,(\oplus/)\,(\oplus/) \cdot \functF\,\return\,\id \\
= & \qquad \{\; \mbox{functors, monad algebras} \;\} \\
  & \foldr\,(\otimes)\,b \cdot \contents_{\functF} \cdot \functF\,\id\,(\oplus/) \\
= & \qquad \{\; \mbox{defining \(f\) in terms of \(\contents\)} \;\} \\
  & f \cdot \functF\,\id\,(\oplus/)
\end{array} \]
\end{answer} \fi

\item (See page~\pageref{use:distr-fusion}.) \label{ex:distr-fusion}
Use fold fusion to prove that
\[ \foldr\,(\otimes)\,b \cdot \functList\,(\oplus/) =
\oplus/ \cdot \functM\,(\foldr\,(\otimes)\,b) \cdot \delta_{\functList}
\]
assuming the distributivity property
$\oplus/ \cdot \functM\,(\otimes) \cdot \mathit{cp} = (\otimes) \cdot (\oplus/)\times(\oplus/)$
of $\mathit{cp}$.
\ifanswers \begin{answer}
For the base case we have
\[ \oplus/\,(\functM\,(\foldr\,(\otimes)\,b)\,(\return\,[\,]))
= \oplus/\,(\return\,(\foldr\,(\otimes)\,b\,[\,]))
= \oplus/\,(\return\,b)
= b
\]
as required. For the inductive step, we have:
\[ \begin{array}{@{}ll}
  & \oplus/ \cdot \functM\,(\foldr\,(\otimes)\,b) \cdot \functM\,(:) \cdot \mathit{cp} \\
= & \qquad \{\; \mbox{functors} \;\} \\
  & \oplus/ \cdot \functM\,(\foldr\,(\otimes)\,b \cdot (:)) \cdot \mathit{cp} \\
= & \qquad \{\; \mbox{evaluation for \(\foldr\)} \;\} \\
  & \oplus/ \cdot \functM\,((\otimes) \cdot \id\times\foldr\,(\otimes)\,b) \cdot \mathit{cp} \\
= & \qquad \{\; \mbox{functors; naturality of \(\mathit{cp}\)} \;\} \\
  & \oplus/ \cdot \functM\,(\otimes) \cdot \mathit{cp} \cdot \functM\,\id\times\functM\,(\foldr\,(\otimes)\,b) \\
= & \qquad \{\; \mbox{distributivity for \(\mathit{cp}\): see below} \;\} \\
  & (\otimes) \cdot (\oplus/)\times(\oplus/) \cdot \functM\,\id\times\functM\,(\foldr\,(\otimes)\,b) \\
= & \qquad \{\; \mbox{functors} \;\} \\
  & (\otimes) \cdot (\oplus/)\times\id \cdot \id\times\functM\,(\oplus/\cdot\foldr\,(\otimes)\,b)
\end{array} \]
\end{answer} \fi

\item (See page~\pageref{use:distr-cp}.) \label{ex:distr-cp}
Prove the distributivity property for cartesian product
\[
\oplus/ \cdot \functM\,(\otimes) \cdot \mathit{cp} = (\otimes) \cdot (\oplus/)\times(\oplus/)
\]
assuming the two distributivity properties 
$\oplus/ \cdot \functM\,(a\otimes) = (a\otimes) \cdot \oplus/$ and
$\oplus/ \cdot \functM\,(\otimes b) = (\otimes b) \cdot \oplus/$
for collections.
\ifanswers \begin{answer}
Writing $\dollar$ for right-associative, loosest-binding application, 
to reduce parentheses, we have:
\[ \begin{array}{@{}ll}
  & \oplus/ \dollar \functM\,(\otimes) \dollar \mathit{cp}\,(x,y) \\
= & \qquad \{\; \mbox{definition of \(\mathit{cp}\)} \;\} \\
  & \oplus/ \dollar \functM\,(\otimes) \dollar \join \dollar \functM\,(\lambda a \mathbin{.} \functM\,(a,)\,y)\,x \\
= & \qquad \{\; \mbox{naturality} \;\} \\
  & \oplus/ \dollar \join \dollar \functM\,(\lambda a \mathbin{.} \functM\,(a\otimes)\,y)\,x \\
= & \qquad \{\; \mbox{\(\oplus/\) is an \(\functM\)-algebra} \;\} \\
  & \oplus/ \dollar \functM\,(\oplus/) \dollar \functM\,(\lambda a \mathbin{.} \functM\,(a\otimes)\,y)\,x \\
= & \qquad \{\; \mbox{functors} \;\} \\
  & \oplus/ \dollar \functM\,(\lambda a \mathbin{.} \oplus/(\functM\,(a\otimes)\,y))\,x \\
= & \qquad \{\; \mbox{distributivity for collections: see below} \;\} \\
  & \oplus/ \dollar \functM\,(\lambda a \mathbin{.} a \otimes (\oplus/\,y))\,x \\
= & \qquad \{\; \mbox{sectioning} \;\} \\
  & \oplus/ \dollar \functM\,(\otimes (\oplus/\,y))\,x \\
= & \qquad \{\; \mbox{distributivity for collections again} \;\} \\
  & (\otimes (\oplus/\,y))\,(\oplus/\,x) \\
= & \qquad \{\; \mbox{sectioning} \;\} \\
  & (\oplus/\,x) \otimes (\oplus/\,y) \\
= & \qquad \{\; \mbox{eta-expansion} \;\} \\
  & (\otimes) \dollar (\oplus/ \times \oplus/) \dollar (x,y) \\
\end{array} \]
\end{answer} \fi

\item (See page~\pageref{use:distr-coll}.) \label{ex:distr-coll}
Prove the two distributivity properties for collections
\[ \begin{array}{@{}lcl}
\oplus/ \cdot \functM\,(a\otimes) &=& (a\otimes) \cdot \oplus/ \\
\oplus/ \cdot \functM\,(\otimes b) &=& (\otimes b) \cdot \oplus/ 
\end{array} \]
by induction over the size of the (finite!) collection,
assuming that binary operator $\otimes$ distributes
over $\oplus$ in the familiar sense 
(that is, $a \otimes (b \oplus c) = (a \otimes b) \oplus (a \otimes c)$).
\ifanswers \begin{answer}
The base cases are for empty and
singleton collections; the case for the empty collection follows from
$e_\oplus$ being a zero of $\otimes$ (that is, $e_\oplus \times a =
e_\oplus = a \times e_\oplus$ for any $a$), and the case for singleton
collections, ie those in the image of $\return$, follows
from the fact that $\oplus/$ is an $\functM$-algebra. The inductive
step is for a collection of the form $u \mplus v$ with $u,v$ both
strictly smaller than the whole (so, if the monad is idempotent, we
should use disjoint union, or at least not the trivial union of a
collection with one of its subcollections); this requires the
distribution of the algebra over choice $\oplus / (u \mplus v) =
(\oplus/u) \oplus (\oplus/v)$, together with the familiar distribution
of $\otimes$ over $\oplus$.
\end{answer} \fi

\end{enumerate}

\end{document}